\documentclass[aps,prd,twocolumn,groupedaddress,nofootinbib]{revtex4}
\usepackage{amsmath,graphicx,amssymb,slashed}
\usepackage[hidelinks]{hyperref}
\usepackage[greek,english]{babel}
\newcommand{\cpi}{\text{\greektext p}}

\newcommand{\iu}{{\mathrm i}}
\newcommand{\E}{{\mathrm e}}

\begin{document}

%\preprint{}

\title{The Quantum Mechanics Swampland}


\author{Aditya Parikh}
%\thanks{}
\affiliation{Center for the Fundamental Laws of Nature, Harvard University, Cambridge, MA, 02138}


%\date{\today}

\begin{abstract}
We investigate non-relativistic quantum mechanical potentials between fermions generated by various classes of QFT operators and evaluate their singularity structure. These potentials can be generated either by four-fermion operators or by the exchange of a scalar or vector mediator coupled via renormalizable or non-renormalizable operators. In the non-relativistic regime, solving the Schr\"odinger equation with these potentials provides an accurate description of the scattering process. This procedure requires providing a set of boundary conditions. We first recapitulate the procedure for setting the boundary conditions by matching the first Born approximation in quantum mechanics to the tree-level QFT approximation. Using this procedure, we show that the potentials are nonsingular, despite the presence of terms proportional to $r^{-3}$ and $\nabla_{i}\nabla_{j}\delta^{3}(\vec{r})$. This surprising feature leads us to propose the \emph{Quantum Mechanics Swampland}, in which the Landscape consists of non-relativistic quantum mechanical potentials that can be UV completed to a QFT, and the Swampland consists of pathological potentials which cannot. We identify preliminary criteria for distinguishing potentials which reside in the Landscape from those that reside in the Swampland. We also consider extensions to potentials in higher dimensions and find that Coulomb potentials are nonsingular in an arbitrary number of spacetime dimensions.
\end{abstract}

%\keywords{}

\maketitle

\section{Introduction}
The general idea of the Swampland program is to isolate low energy effective QFTs which can be consistently coupled to quantum gravity in the UV. From a low energy perspective, we are free to evaluate an infinitude of QFTs with various types of properties, such as $SU(N)$ gauge theories with any $N$ and arbitrarily small gauge couplings. On the other hand, many of these QFTs develop inconsistencies when we couple to quantum gravity. Consistency then becomes a powerful tool for determining what classes of QFTs are viable. These various consistency conditions are formulated into the Swampland conjectures. QFTs which satisfy these conjectures live in the Landscape whereas theories which violate them live in the Swampland~\cite{Vafa:2005ui,Brennan:2017rbf,Palti:2019pca}. 

In the same spirit, we can apply this idea to the study of quantum mechanical potentials which are a low energy effective description of a scattering process in a QFT. We are free to consider any potential we want, but if the eventual goal is to obtain an underlying QFT description of the dynamics we are trying to model, then consistency again serves to limit the space of viable potentials. This analogous relationship motivates the existence of the \emph{Quantum Mechanics Swampland} where potentials that can be consistently derived from a QFT belong in the Landscape and potentials which cannot belong in the Swampland. 
We emphasize that the Quantum Mechanics Swampland we propose is \emph{not} the original Swampland, but there exist parallels in both the philosophy and ultimate goal of the respective programs, and as such it is useful to invoke similar terminology. The Swampland program in QFT aims to identify the boundary between apparently consistent QFTs which can be UV completed into a theory of quantum gravity and those which cannot. Likewise, the Quantum Mechanics Swampland aims to delineate the boundary between apparently consistent quantum mechanical potentials which can be UV completed into a QFT and those which cannot. The Swampland program achieves this through a series of conjectures outlining properties which low energy QFTs are expected to have. QFTs which satisfy these criteria are ``good" theories residing in the Landscape which might have a chance of describing the world around us at all scales. In a similar fashion, in this paper we propose a preliminary criteria that quantum mechanical potentials must satisfy so that they are ``good" potentials residing in the Quantum Mechanics Landscape. Since these potentials can be UV completed into a QFT, they have a chance of describing the phenomena we see around us in the non-relativistic and the relativistic regime. In contrast, potentials which reside in the Quantum Mechanics Swampland can never be UV completed to a QFT, so regardless of how good of an empirical fit they might be, the underlying microphysics will not be amenable to a QFT description.

%Similar to the Swampland program, the goal will be to outline criteria which will help delineate the boundary between the Swampland and the Landscape. 

Exploring the relationship between quantum mechanical potentials and field theory operators is certainly not new. It was originally studied in the context of Bethe-Salpeter equations in~\cite{Bastai:1963tsbs}. The authors concluded that super-renormalizable operators yielded regular potentials $V_{r}(r)$, renormalizable operators yielded transition potentials $V_{t}(r)$, and non-renormalizable operators yielded singular potentials $V_{s}(r)$ (see~\cite{Frank:1971a} for a review of singular potentials). These potentials satisfied the following conditions at the origin. Here, $C$ is a finite constant.
\begin{equation}
\begin{split}
    &\lim_{r\to 0}|r^{2}V_{r}(r)| = 0 \\
    &\lim_{r\to 0}|r^{2}V_{t}(r)| = C \\
    &\lim_{r\to 0}|r^{2}V_{s}(r)| = \infty
\end{split}
\end{equation}
Later, Lepage and Caswell developed a non-relativistic effective field theory approach to this problem in~\cite{Caswell:1985ui, Lepage:1997cs}. Although they didn't seek to address the relationship between field theory operators and singular potentials, their work simplified the analysis of many non-relativistic phenomena such as low energy scattering due to the new toolkit they introduced which leveraged the non-relativistic nature of the problem. More recently, we developed a novel matching procedure in~\cite{Agrawal:2020lea}, where we rely on a relativistic field theory description at short distances and a non-relativistic quantum mechanical description at large distances. The matching is performed at the Compton radius of the scattered particle, which is the natural scale where relativistic effects start becoming important. In light of this new approach, it is worth readdressing the question of what classes of field theory operators lead to singular potentials. As we will see, our conclusions about this classification differ from~\cite{Bastai:1963tsbs}. 

In~\cite{Agrawal:2020lea}, we showed that at short distances, there was a match between the tree-level relativistic field theory description and the first Born approximation using the corresponding quantum mechanical potential.
This underlying correspondence between QFT and quantum mechanics naturally leads to the following criteria for determining which potentials live in the Landscape and the Swampland:

\begin{center}
\begin{minipage}{20em}
%\begin{itemize}
The Quantum Mechanics Landscape consists of quantum mechanical potentials arising from a well-defined QFT scattering process. As a consequence of this definition, from the IR perspective, a quantum mechanical potential belongs in the Quantum Mechanics Swampland if it is singular and results in a divergent first Born approximation.
%\item From the IR perspective, a quantum mechanical potential belongs in the Swampland if it is singular and gives rise to a divergent first Born approximation.
%\item From the UV perspective, a well-defined tree-level QFT scattering process always gives rise to non-singular potentials that live in the Landscape.
%\end{itemize}
\end{minipage}
\end{center}

The analysis of low energy scattering is ubiquitous in various branches of physics. Non-relativistic nucleon-nucleon scattering~\cite{Naghdi:2007ek,Epelbaum:2008ga} was a helpful tool in understanding the strong force. Non-relativistic scattering is also of central importance in determining properties of dark matter. Scattering processes include dark matter direct detection~\cite{Agrawal:2010fh,Fan:2010gt,Freytsis:2010ne,Fitzpatrick:2012ix,Bishara:2016hek} and dark matter scattering and annihilation in galaxies~\cite{Hisano:2002fk,Hisano:2003ec,Hisano:2004ds,ArkaniHamed:2008qn,Buckley:2009in,Loeb:2010gj,Tulin:2013teo,Bellazzini:2013foa,Kaplinghat:2015aga,Blum:2016nrz,Agrawal:2020lea}. Furthermore, cross sections can be nonperturbatively enhanced via the Sommerfeld effect~\cite{sommerfeld1931beugung} in the non-relativistic regime. Analysis of the Sommerfeld effect, which was carried out in~\cite{Agrawal:2020lea} in the context of self-interacting dark matter models, even showed that the potential generated by pseudoscalar exchange does not lead to any enhancement. This was a new result showing that the matching procedure was sufficient and there was no need to renormalize, as had previously been suggested in the literature. In addition, we also showed that the operator accompanying the potential was critical in reproducing the physics, and it was incorrect to approximate it as a simple $1/r^{3}$ central potential. 
With such a wide range of applications, identifying which potentials reside in the Landscape and which ones reside in the Swampland becomes crucial for ensuring that the empirical description of a low energy phenomenon can be consistently completed into a QFT. This classification is also an important tool for effective field theorists tasked with building a theoretical description underlying these low energy processes, since it provides theoretical input on what classes of potentials are viable.

%Furthermore, this classification scheme and matching procedure can lead to interesting results even at tree-level. 

%For example, it is important in dark matter physics when considering processes such as dark matter direct detection~\cite{Agrawal:2010fh,Fan:2010gt,Freytsis:2010ne,Fitzpatrick:2012ix,Bishara:2016hek}, Sommerfeld enhancement~\cite{sommerfeld1931beugung}, and dark matter scattering and annihilation in galaxies~\cite{Hisano:2002fk,Hisano:2003ec,Hisano:2004ds,ArkaniHamed:2008qn,Buckley:2009in,Loeb:2010gj,Tulin:2013teo,Bellazzini:2013foa,Kaplinghat:2015aga,Blum:2016nrz,Agrawal:2020lea}, as well as nuclear physics where we analyze nucleon-nucleon scattering~\cite{Naghdi:2007ek,Epelbaum:2008ga}. Identifying which potentials reside in the Landscape and which ones reside in the Swampland becomes crucial for ensuring that the empirical description of a low energy phenomenon can be consistently completed into a QFT. This classification is also an important tool for effective field theorists tasked with building a theoretical description underlying these low energy processes, since it provides theoretical input on what classes or potentials are viable.

To begin exploring the Landscape, in this paper we will focus on perturbative QFTs. We will derive the potential experienced by fermions coupled in a variety of ways subject to a tree-level matching between QFT and quantum mechanics. Our focus will be on fermion-fermion scattering because each particle participating in the scattering has intrinsic spin which leads to a larger variety of possible operator structures in the non-relativistic potential. We briefly comment on the scalar-scalar and scalar-fermion scattering cases which work analogously, in Appendix~\ref{sec:scalars}. We will begin by reviewing how to set up the initial conditions for the scattering process of interest in Section~\ref{sec:scattering_setup}. Then we will start our investigation of potentials. Starting from an underlying QFT, we will study the potentials generated from tree-level approximations to the field theory scattering process. This can be generated by renormalizable interactions coupling fermions and mediators, which we study in Section~\ref{sec:rel_ops}, or non-renormalizable interactions, which we study in Section~\ref{sec:nonrel_ops}. We relegate some detailed calculations to Appendix~\ref{sec:delta_fcn_derivatives}. Having derived all of these potentials, we will be in a position to critically address which potentials are truly singular or not. This will conclude our study of tree-level potentials in 3 + 1-dimensions. In Section~\ref{sec:Higher_Dimensions}, we extend our tree-level results to higher dimensions. We offer concluding remarks in Section~\ref{sec:conclusions}.

\section{Setting Up the Scattering Calculation}
\label{sec:scattering_setup}
In this section we will begin by discussing the derivation of potentials and then review the procedure we formulated in~\cite{Agrawal:2020lea} for setting the boundary conditions.
Given a set of interactions in a perturbative QFT, we can write down the tree-level QFT amplitude for a particular scattering process.
In non-relativistic quantum mechanics, the Born approximation to the scattering amplitude is given by
\begin{equation}
\langle \vec{p}_{\rm f} | \iu \mathcal{T} | \vec{p}_{\rm i} \rangle = -\iu \widetilde{V}(\vec{q})(2\cpi)\delta(E_{p_{\rm f}}-E_{p_{\rm i}}), \quad \quad \vec{q} = \vec{p}_{\rm f} - \vec{p}_{\rm i}.
\end{equation}
If the relativistic corrections are small, then we can match the non-relativistic limit of the QFT amplitude to the quantum mechanical amplitude. This gives us the following expression for the Fourier transform of the quantum mechanical potential, $\Tilde{V}(\vec{q})$, in terms of the QFT amplitude
\begin{equation}
    \Tilde{V}(\vec{q}) = -\frac{1}{2E_{p_{f}}}\frac{1}{2E_{p_{i}}}M.
\end{equation}
Relativistic and non-relativistic single particle states have a relative factor of $2E_{p}$ in their normalizations which is taken into account by the prefactor. The quantum mechanics calculation occurs in the center-of-mass frame. This effectively reduces it to a single particle problem, and hence provides only a single factor of $2E_{p}$ for the initial and final state. Once we have $\Tilde{V}(\vec{q})$, we can Fourier transform with respect to $\vec{q}$ to compute the potential $V(\vec{r})$ in real space. We follow this procedure to arrive at the results in Section~\ref{sec:Tree_Level}. 

\begin{figure}
\includegraphics[width=0.5\textwidth]{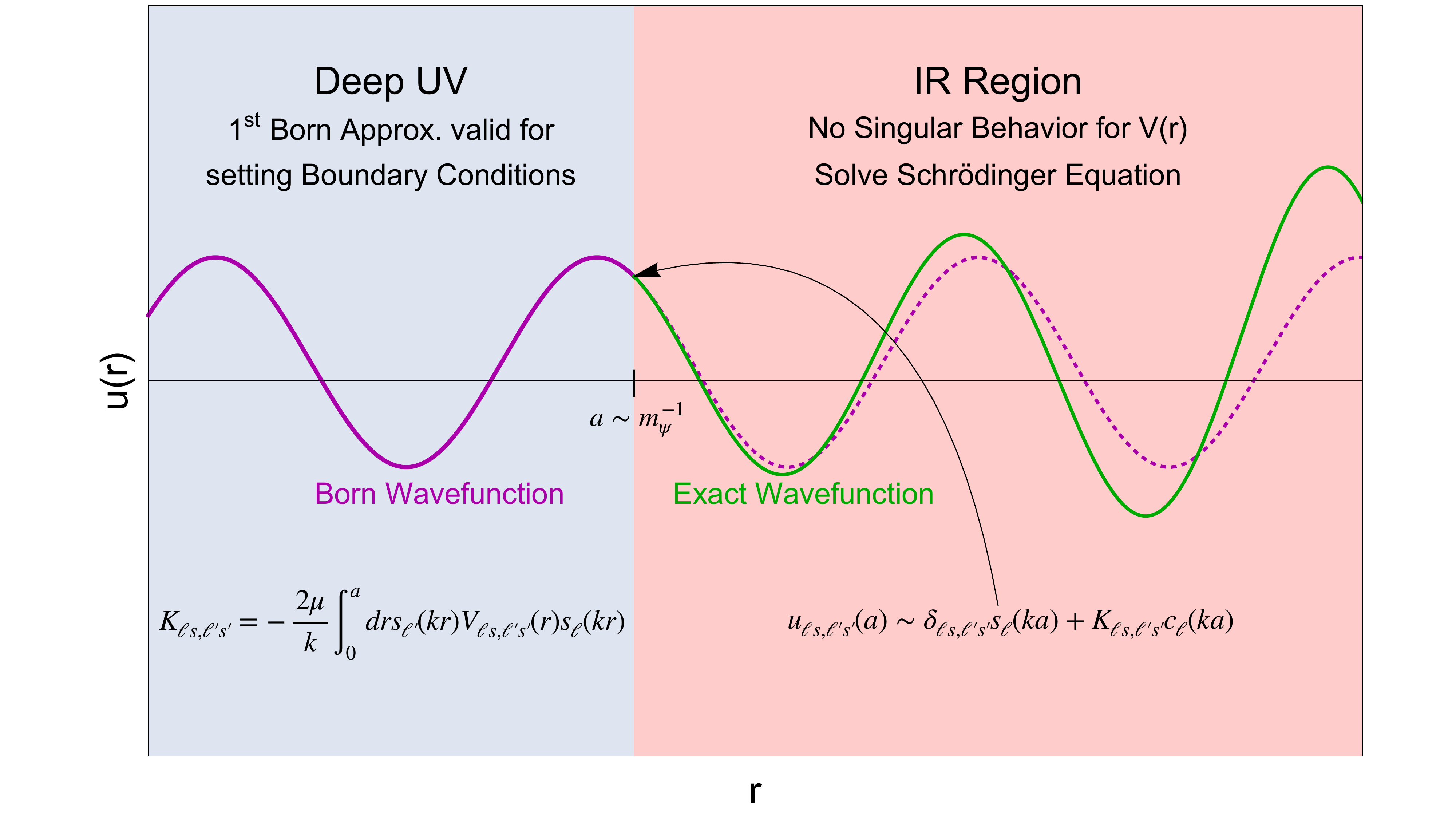}
\caption{A schematic for the matching procedure adapted from \cite{Agrawal:2020lea}. We use the Born approximation to set the boundary conditions since we expect it to approximate the scattering process well in the deep UV. $r > a$ probes the IR of our theory where we expect the quantum mechanical potential to be a good description of our system. Since we have excised the origin, we no longer have to worry about the (potentially) singular nature of the potential while solving the Schr\"odinger equation. The exact wavefunction, which is the solution to the Schr\"odinger equation, can deviate from the Born wavefunction and these deformations are of physical significance, as evidenced by the analysis of Sommerfeld enhancement.\label{fig:IC_Schematic}}
\end{figure}

The observable of interest for a low energy scattering process is typically a cross section. In the non-relativistic regime, there can be significant deviations from a tree-level QFT approximation, for example from non-perturbative Sommerfeld enhancement~\cite{sommerfeld1931beugung}. A convenient way of accounting for these effects is to map the problem to an effective quantum mechanical problem, solving the Schr\"odinger equation and extracting the scattering matrix elements. To solve the Schr\"odinger equation, we need to supply appropriate boundary conditions, in addition to the potential. The novel insight from~\cite{Agrawal:2020lea} was that the physics occurring on short distance scales influences the wavefunction, which in turn affects the boundary conditions. To achieve this separation of scales, we split the region and match at the Compton radius of the scattered particle which is given by $a \sim \mathcal{O}(m_{\psi}^{-1})$. The Compton radius is the natural matching scale because for $r < a$, the particles start becoming relativistic with momenta of $\mathcal{O}(m_{\psi})$. The quantum mechanical potential is a non-relativistic effective description of the scattering process, so at these scales, we should not expect this description to hold and must resort to the underlying relativistic QFT description. In~\cite{Agrawal:2020lea}, it was shown numerically that the QFT tree-level amplitude matched with the first Born approximation in quantum mechanics for $r < a$. We showed this for a variety of potentials, and also showed that this matching was robust to variations in the matching scale as long as it was $\mathcal{O}(m_{\psi}^{-1})$. Establishing this correspondence allows us to compute the boundary conditions at $r = a$, accounting for the effects of the short distance physics. For $r > a$, we are in the non-relativistic regime where quantum mechanics is an appropriate description of the physics. In this region, we can solve the Schr\"odinger equation and extract scattering matrix elements which incorporate long distance non-perturbative effects. This procedure ensures a separation of scales and that the appropriate description is used in the respective regimes. We show this schematically in Figure~\ref{fig:IC_Schematic} and proceed to discuss how to implement this procedure in more detail.

%The novel insight from the procedure developed in~\cite{Agrawal:2020lea} was that for some region $r < a$, the QFT tree-level amplitude is matched with the first Born approximation in quantum mechanics.\footnote{This matching was shown numerically in~\cite{Agrawal:2020lea} for a variety of potentials.} 
%For scattering processes, the matching radius is set by the Compton radius of the fermion which is given by $a \sim \mathcal{O}(m_{\psi}^{-1})$. 
%The Compton radius sets the natural matching scale because for $r < a$, the particles start becoming relativistic since they have momenta of $\mathcal{O}(m_{\psi})$. 
%The quantum mechanical potential is a non-relativistic effective description of the scattering process. So, at these scales, we should not expect the non-relativistic description to hold and must resort to the underlying relativistic QFT description. 
%Performing this matching allows us to calculate the boundary conditions at $r = a$, and the Schr\"odinger equation can subsequently be solved. 

If we assume $rV(r) \to 0$ as $r \to \infty$, then asymptotically, the wavefunction is a solution of the free particle Schr\"odinger equation. $s_{\ell}$ and $c_{\ell}$ are solutions to the radial equation. They are given by
\begin{equation}
    s_{\ell}(kr) \equiv krj_{\ell}(kr) \quad \quad c_{\ell}(kr) \equiv -kry_{\ell}(kr)
\end{equation}
where the $j_{\ell}(kr)$ and $y_{\ell}(kr)$ are spherical Bessel functions and $\ell$ is the angular momentum of the corresponding partial wave. These solutions form a basis that the asymptotic solution to the full Schr\"odinger equation can be decomposed on. The decomposition is given by
\begin{equation}
    u_{\ell s,\ell's'}(r) \sim \delta_{\ell s,\ell's'}s_{\ell}(kr) + K_{\ell s,\ell's'}c_{\ell}(kr)
\label{eqn:asymptotic_wavefcn}
\end{equation}
$u_{\ell s,\ell's'}$ is the reduced radial wavefunction. $K_{\ell s,\ell's'}$ is a matrix which generalizes partial wave phase shifts to account for interactions coupling various partial waves. As we will see shortly, this will be relevant for potentials such as those induced by pseudoscalar exchange. In this case, the differential equations are coupled, so that the potential $V(r)$ is now a matrix $V_{\ell s, \ell' s'}(r)$.

For a perturbative QFT, the tree-level QFT approximation of a particular matrix element will be faithfully reproduced by the first Born approximation which is given by
\begin{equation}
K_{\ell s,\ell's'}=\frac{-2\mu}{k}\int_{0}^{a} dr s_{\ell'}(kr)V_{\ell s,\ell's'}(r)s_{\ell}(kr)
\label{eqn:tan_delta_l}
\end{equation}
In an operator language, the integrand of Equation~\ref{eqn:tan_delta_l} can be understood as
\begin{equation}
    s_{\ell'}(kr)V_{\ell s,\ell's'}(r)s_{\ell}(kr) \propto \langle \ell' | \hat{V}_{\ell s,\ell's'} | \ell \rangle
\end{equation}
where $\hat{V}_{\ell s,\ell's'}$ is now an operator and $| \ell \rangle$ are states with angular momentum $\ell$.  The boundary conditions at $r = a$ are then given by
\begin{equation}
    u_{\ell s,\ell's'}(a) \sim \delta_{\ell s,\ell's'}s_{\ell}(ka) + K_{\ell s,\ell's'}c_{\ell}(ka)
\label{eqn:bc_wavefcn}
\end{equation}

Using the method we described above, we now have a clear analytical diagnostic for evaluating whether a potential is singular or not:
\vspace{1em}
\begin{center}
\begin{minipage}{20em}
\textbf{Diagnostic:} If the integral in Equation~\ref{eqn:tan_delta_l} diverges for \emph{any} combination of incoming and outgoing states, then the potential is singular.
\newline
\newline
\newline
\end{minipage}
\end{center}
Since we are working in the non-relativistic regime, the argument $kr$ is always small over the interval of integration. For small arguments, we expand the spherical Bessel function and find that $s_{\ell}(kr) \sim (kr)^{\ell+1}$. Setting $\ell = \ell' = 0$ in Equation~\ref{eqn:tan_delta_l} and using the small argument expansion of $s_{\ell}$, we recover the familiar fact that if a potential diverges faster than $r^{-2}$ as $r \rightarrow 0$, the potential might be singular. We now derive tree-level potentials for various interactions and analyze whether they are singular or not using the analytic diagnostic discussed above.
\newline

\section{Tree-Level Potentials}
\label{sec:Tree_Level}
In this section, we will compute the tree-level potentials experienced by fermions coupled via various interactions. By tree-level potential here we mean that the amplitude in the QFT is the tree-level amplitude. For concreteness, we will study the process $\psi_{1}\psi_{2}\rightarrow\psi_{1}\psi_{2}$ where $\psi_{i}$ are spin-1/2 fermions. We will consider cases with and without a mediator.

\subsection{Renormalizable Interactions}
\label{sec:rel_ops}
We begin by considering potentials generated by renormalizable operators in the QFT. For concreteness we will consider the following operators
\begin{equation}
\begin{split}
&\mathcal{L}_{\text{\bf s}} = \lambda\phi\overline{\psi}\psi \quad \mathcal{L}_{\text{\bf ps}} = \iu\lambda\phi\overline{\psi}\gamma^{5}\psi \\
&\mathcal{L}_{\text{\bf v}} = \lambda \phi^{\mu}\overline{\psi} \gamma_{\mu}\psi \quad \mathcal{L}_{\text{\bf av}} = \lambda\phi^{\mu}\overline{\psi}\gamma_{\mu}\gamma^{5}\psi
\label{eqn:interactions}
\end{split}
\end{equation} 
Here the subscripts $\bf s$, $\bf ps$, $\bf v$, and $\bf av$ denote scalar, pseudoscalar, vector and axial vector respectively. To encompass all possible pairings of interactions, we consider two fermion species $\psi_{1}$ and $\psi_{2}$ each of which is independently subject to one of the interactions in Equation \ref{eqn:interactions}. At tree-level, the process $\psi_{1}\psi_{2}\rightarrow\psi_{1}\psi_{2}$ only has a t-channel Feynman diagram contributing to it.\footnote{Processes like $\psi\bar{\psi}\rightarrow\psi\bar{\psi}$ also have a contribution from an s-channel Feynman diagram. This gives rise to a $\delta^{3}(\vec{r})$ contact interaction in the non-relativistic limit. It can explicitly be shown, by computing Equation~\ref{eqn:tan_delta_l}, that $\delta^{3}(\vec{r})$ gives a finite nonsingular result for all possible values of $\ell$ and $\ell'$.} From the tree-level amplitude, we obtain the following potentials
\begin{widetext}
\begin{equation}
V_{\text{\bf s,s}}(r) = -\frac{\lambda_{1}\lambda_{2}}{4\cpi r}\E^{-m_{\phi}r}
\end{equation}

\begin{equation}
V_{\text{\bf ps,ps}}(r) = \frac{\lambda_{1}\lambda_{2}}{4\cpi m_{1}m_{2}}\E^{-m_{\phi}r}\Bigg[\Big(\frac{m_{\phi}^{2}}{3r} - \frac{4\cpi
\delta^{3}(\vec{r})}{3}\Big)\vec{S}_{1}\cdot \vec{S}_{2} + \frac{\mathcal{O}_{T}}{r^{3}}\Big(1 + m_{\phi}r + \frac{m^{2}_{\phi}r^{2}}{3}\Big)\Bigg]
\label{pseudoscalar_potential_equation}
\end{equation}

\begin{equation}
V_{\text{\bf v,v}}(r) = \frac{\lambda_{1}\lambda_{2}}{4\cpi r}\E^{-m_{\phi}r}
\end{equation}

\begin{equation}
\label{axial_vector_potential_equation}
V_{\text{\bf av,av}}(r) = -\frac{\lambda_{1}\lambda_{2}}{\cpi r}\E^{-m_{\phi}r}\vec{S}_{1}\cdot \vec{S}_{2} + \frac{4m_{1}m_{2}}{m_{\phi}^{2}}V_{\text{\bf ps,ps}}
\end{equation}

\begin{equation}
V_{\text{\bf s,ps}}(r) = \frac{\lambda_{1}\lambda_{2}}{4\cpi m_{2}}\frac{1 + m_{\phi} r}{r^{2}}\E^{-m_{\phi}r}\vec{S}_{2}\cdot\hat{r}
\end{equation}

\begin{equation}
V_{\text{\bf v,av}}(r) = \frac{\lambda_{1}\lambda_{2}}{2\cpi}\frac{\E^{-m_{\phi}r}}{r}\Bigg[\frac{1 + m_{\phi}r}{m_{1}r}(\vec{S}_{1}\times \vec{S}_{2})\cdot\hat{r} + \vec{S}_{2}\cdot\Bigg(\frac{\vec{p}_{2}}{m_{2}} - \frac{\vec{p}_{1}}{m_{1}}\Bigg) + \frac{\iu (m_{1} + m_{2})}{2m_{1}m_{2}}\frac{1 + m_{\phi}r}{r}\vec{S}_{2}\cdot\hat{r}\Bigg]
\end{equation}
\end{widetext}
\pagebreak

The potentials we derive above agree with previous results~\cite{Moody:1984ba,Dobrescu:2006au,Fan:2010gt,Fitzpatrick:2012ix,Bellazzini:2013foa,Daido:2017hsl,Fadeev:2018rfl}\footnote{$V_{\text{\bf s,ps}}(r)$ and $V_{\text{\bf v,av}}(r)$ differ by an overall sign relative to the results in~\cite{Fadeev:2018rfl}. This discrepancy arises because we choose $\psi_{1}$ to have a scalar (or vector) interaction and $\psi_{2}$ to have a pseudoscalar (or axial vector) interaction while they consider the opposite scenario.}. The subscripts for $V$ indicate the type of interaction vertices present in the Feynman diagram. The first subscript denotes the $\psi_{1}$ coupling and the second subscript denotes the $\psi_{2}$ coupling. Here, $m_{i}$ denotes the mass of fermion $\psi_{i}$ and $\vec{S}_{i}$ its spin. We also define the operator $\mathcal{O}_{T}$ as
\begin{equation}
    \mathcal{O}_{T} = 3(\vec{S}_{1}\cdot\hat{r})(\vec{S}_{2}\cdot\hat{r}) - \vec{S}_{1}\cdot \vec{S}_{2}
\end{equation}
We note that the only potential which has terms diverging faster than $r^{-2}$ near the origin is $V_{\text{\bf ps,ps}}$ (and in turn $V_{\text{\bf av,av}}$). The $\mathcal{O}_{T}$ term is potentially problematic since it has a piece diverging as $r^{-3}$. In particular, the first Born approximation is well-behaved for every combination of states besides potentially the $\ell = \ell' = 0$ case. While this scenario could be divergent, in~\cite{Agrawal:2020lea} we showed that $\langle \ell' = 0 | \mathcal{O}_{T} | \ell = 0 \rangle = 0$. So, the operator structure in the potential prevents the divergence from arising and $V_{\text{\bf ps,ps}}$ is not singular. Therefore, all of the potentials we consider in this section are nonsingular.
\newline

\subsection{Non-Renormalizable Interactions}
\label{sec:nonrel_ops}
The case of potentials arising from non-renormalizable interactions factors into two scenarios: ones with a mediator and ones without. Examples of the former include the fermions coupled to a field strength tensor or derivatively coupled to a Goldstone while examples of the latter include four-fermion operators.

We begin by analyzing the Goldstone and tensor couplings
\begin{equation}
    \mathcal{L}_{\text{Goldstone}} = \frac{1}{\Lambda}\overline{\psi}\gamma^{\mu}\gamma^{5}\psi\partial_{\mu}\phi \quad \quad \mathcal{L}_{\text{tensor}} = \frac{1}{2\Lambda}\overline{\psi}\sigma^{\mu\nu}\psi F_{\mu\nu}
\label{eqn:nonrel_ops}
\end{equation}
These couplings are examples of scenarios where we have a scalar or vector mediator but the coupling is non-renormalizable. Here we define $\sigma^{\mu\nu} = \frac{\iu}{4}[\gamma^{\mu},\gamma^{\nu}]$ and $F_{\mu\nu} = \partial_{\mu}\phi_{\nu} - \partial_{\nu}\phi_{\mu}$.

From the tree-level amplitude, we obtain the following potentials
\begin{widetext}
\begin{equation}
    V_{\text{Goldstone}}(r) = \frac{\E^{-m_{\phi}r}}{\cpi \Lambda^{2}}\Bigg[\Big(\frac{m_{\phi}^{2}}{3r} - \frac{4\cpi\delta^{3}(\vec{r})}{3}\Big)\vec{S}_{1}\cdot \vec{S}_{2} + \frac{\mathcal{O}_{T}}{r^{3}}\Big(1 + m_{\phi}r + \frac{m^{2}_{\phi}r^{2}}{3}\Big)\Bigg]
\label{Goldstone_potential_equation}
\end{equation}

\begin{equation}
    V_{\text{tensor}}(r) =  \frac{\E^{-m_{\phi}r}}{4\cpi \Lambda^{2}}\Bigg[ \frac{\mathcal{O}_{T}}{r^{3}}\Big(1 + m_{\phi}r + \frac{m^{2}_{\phi}r^{2}}{3}\Big) - \Big(\frac{2m_{\phi}^{2}}{3r} + \frac{4\cpi\delta^{3}(\vec{r})}{3}\Big)\vec{S}_{1}\cdot \vec{S}_{2}\Bigg] + \frac{\delta^{3}(\vec{r})}{\Lambda^{2}}\vec{S}_{1}\cdot\vec{S}_{2}
\end{equation}
\end{widetext}

We can also consider tensor couplings of the form $\iu\overline{\psi}\sigma^{\mu\nu}\gamma^{5}\psi F_{\mu\nu}$. Together, these two tensor couplings are the magnetic and electric dipole interactions, respectively. These two interactions are related via $2\iu\sigma^{\sigma\rho} = \epsilon^{\mu\nu\sigma\rho}\sigma_{\mu\nu}\gamma^{5}$. The potential generated by this interaction has an additional $\epsilon$ tensor inserted at each vertex, but this does not change the overall radial dependence of the potential. The difference between these two interactions manifests when we consider monopole-dipole type couplings. Particles with an intrinsic electric dipole moment experience enhancements in the non-relativistic regime when scattering off a charged particle, whereas particles with intrinsic magnetic dipole moments do not. Understanding the nature of this enhancement warrants solving the Schr\"odinger equation with the appropriate potential. For the purpose of our study, we are only interested in the structure of the potential and whether it gives rise to a finite first Born approximation. It can be shown that this is the case for both tensor potentials and the Goldstone potential, so all these potentials reside in the Landscape.

Next, we consider the scenario where we don't have a mediator. Scenarios like this arise when a heavy mediator has been integrated out leaving effective four fermion operators. The most general four fermion operator we can write down has the form
\begin{equation}
    \frac{\lambda}{\Lambda^{2}}\overline{\psi}_{1}\Gamma_{1}\psi_{1}\overline{\psi}_{2}\Gamma_{2}\psi_{2}
\label{eqn:four_fermi_ops}
\end{equation}
In Table~\ref{tab:nonrel_potentials}, we tabulate the leading non-relativistic potentials in position space for various four fermion operators, which agree with the results in~\cite{Fan:2010gt}.

\begin{table*}
\caption{Leading non-relativistic potentials generated for various effective four fermion operators.\label{tab:nonrel_potentials}}
\begin{tabular}{|c|c|}
\hline
Effective Operator & Position Space Potential \\
\hline
$\frac{\lambda}{\Lambda^{2}}\overline{\psi}_{1}\psi_{1}\overline{\psi}_{2}\psi_{2}$ & $\frac{\lambda}{\Lambda^{2}}\delta^{3}(\vec{r})$ \\
 & \\
$\frac{\lambda}{\Lambda^{2}}\overline{\psi}_{1}\gamma^{5}\psi_{1}\overline{\psi}_{2}\gamma^{5}\psi_{2}$ & $\frac{\lambda}{m_{1}m_{2}\Lambda^{2}}(\vec{S}_{1}\cdot\vec{\nabla})(\vec{S}_{2}\cdot\vec{\nabla})\delta^{3}(\vec{r})$ \\
 & \\     
$\frac{\iu\lambda}{\Lambda^{2}}\overline{\psi}_{1}\psi_{1}\overline{\psi}_{2}\gamma^{5}\psi_{2}$ & $\frac{\lambda}{m_{2}\Lambda^{2}}(\vec{S}_{2}\cdot\vec{\nabla})\delta^{3}(\vec{r})$ \\
 & \\        
$\frac{\lambda}{\Lambda^{2}}\overline{\psi}_{1}\gamma^{\mu}\psi_{1}\overline{\psi}_{2}\gamma_{\mu}\psi_{2}$ & $\frac{\lambda}{\Lambda^{2}}\delta^{3}(\vec{r})$ \\
 & \\       
\quad $\frac{\lambda}{\Lambda^{2}}\overline{\psi}_{1}\gamma^{5}\gamma^{\mu}\psi_{1}\overline{\psi}_{2}\gamma^{5}\gamma_{\mu}\psi_{2}$ \quad & $-\frac{4\lambda}{\Lambda^{2}}\vec{S}_{1}\cdot \vec{S}_{2}\delta^{3}(\vec{r})$ \\
 & \\        
$\frac{\lambda}{\Lambda^{2}}\overline{\psi}_{1}\gamma^{\mu}\psi_{1}\overline{\psi}_{2}\gamma^{5}\gamma_{\mu}\psi_{2}$ & \quad $\frac{\lambda}{\Lambda^{2}}\Bigg(2\vec{S}_{2}\cdot\Big(\frac{\vec{p}_{2}}{m_{2}} - \frac{\vec{p}_{1}}{m_{1}}\Big) - \iu \frac{m_{1}+m_{2}}{m_{1}m_{2}}(\vec{S}_{2}\cdot\vec{\nabla}) + \frac{2}{m_{1}}(\vec{S}_{2}\times\vec{S}_{1})\cdot\vec{\nabla}\Bigg)\delta^{3}(\vec{r})$ \quad \\
\hline
\end{tabular}
\end{table*}

The Goldstone and tensor potentials both have potentially problematic terms which are accompanied by $\mathcal{O}_{T}$ preventing the singularity from arising. The potentials from four fermion operators are all accompanied by $\delta^{3}(\vec{r})$, as seen in Table~\ref{tab:nonrel_potentials}. For a simple delta function, the integrand of Equation~\ref{eqn:tan_delta_l} scales as $\delta(r)r^{\ell+\ell'}$ which is finite for $\ell = \ell' = 0$ and zero otherwise. Potentials with derivatives might seem problematic at first, but we show in detail in Appendix~\ref{sec:delta_fcn_derivatives} that they also produce finite well-behaved first Born approximations. So we see that any tree-level potential arising from a QFT is not singular. In the next section, we consider generalizations to arbitrary spatial dimensions.

\section{Extension to Arbitrary Dimensions}
\label{sec:Higher_Dimensions}
We begin by considering the Schr\"odinger equation for a particle that is not subject to any potential, in $d$ spatial dimensions~\cite{Esposito:1998kr,Dong:1999ib,Gonul:2001xc,Caruso_2013}. This is given by
\begin{equation}
    -\frac{1}{2\mu}\nabla_{d}^{2}\Psi (r) = E\Psi (r) \qquad \nabla_{d}^{2} = \partial_{r}^{2} + \frac{d-1}{r}\partial_{r} + \frac{1}{r^{2}}\Omega^{2}
\end{equation}
$\partial_{r}$ represents derivatives with respect to the radial coordinate, $\Omega^{2}$ is the Laplacian on the ($d-1$)-sphere and $\mu$ is the reduced mass. We can decompose $\Psi(r)$ into the product of a radial function $R(r)$ and Gegenbauer polynomials, which are eigenfunctions of $\Omega^{2}$. The Gegenbauer polynomials are a generalization of the spherical harmonics to higher dimensions. We notice that for $d = 3$, we recover the familiar eigenvalue of $\ell(\ell +1)$ for the angular momentum term. Analogous to the 3 + 1-dimensional case, we obtain a radial equation for every harmonic. 
\begin{equation}
    \partial_{r}^{2}R + \frac{d-1}{r}\partial_{r}R -\frac{\ell(\ell+d-2)}{r^{2}}R = -k^{2}R
\end{equation}
To cancel the term with the first derivative of $R(r)$, we can set $u(r) = r^{(d-1)/2}R(r)$. This simplifies the radial equation to
\begin{equation}
    \partial_{r}^{2}u + \bigg[k^{2} - \frac{j(j+1)}{r^{2}}\bigg]u = 0 \qquad j = \ell + \frac{d-3}{2}
\label{eqn:arb_dim_radial}
\end{equation}
The radial free particle solutions are denoted $s_{j}$ and $c_{j}$ and are given in terms of spherical Bessel functions as before, with the only difference being that the order is now dimension dependent. 
\begin{equation}
    s_{j}(kr) = krj_{j}(kr) \qquad c_{j} = -kry_{j}(kr)
\end{equation}
As a test case, we will consider the Coulomb potential which in $d$ spatial dimensions is given by $V(r) = \alpha/r^{d-2}$. For $d > 4$, this potential diverges faster than $r^{-2}$ and there is no accompanying operator structure that can give vanishing matrix elements between potentially problematic states. Naively, this potential appears to be problematic. To test whether this potential is singular, we compute the first Born approximation using Equation~\ref{eqn:tan_delta_l}
\begin{equation}
\begin{split}
    K_{js,j's'} &= \frac{-2\mu}{k}\int_{0}^{a}dr s_{j'}(kr)V_{js,j's'}(r)s_{j}(kr) \\ &\approx \frac{-2\alpha\mu}{k}\int_{0}^{a}dr r^{j'+1}r^{2-d}r^{j+1} \\
    &\approx \frac{-2\alpha\mu}{k}\int_{0}^{a}dr r^{\ell'+\ell+1}
\end{split}
\end{equation}
The most divergent case corresponds to $\ell = \ell' = 0$, which we explicitly see is finite and nonsingular. This is a clear indication that potentials diverging faster than $r^{-2}$ should not be the sole diagnostic for evaluating whether they are singular or not. Furthermore, the dependence on dimension drops out. This indicates that Coulomb potentials, which have well-defined QFT descriptions in any number of dimensions, always reside in the Landscape, and our proposed diagnostic for testing which potential resides in the Landscape works. By extending our analysis to an arbitrary number of dimensions, we see that our earlier results were not just an artifact of working in 3 + 1-dimensions. It also indicates that there exists a deeper connection between QFT and quantum mechanics, where a tree-level perturbative QFT will always produce a nonsingular potential. These results clearly support the existence of the Quantum Mechanics Swampland as well as our criteria for distinguishing which quantum mechanical potentials reside in this Landscape.

\section{Conclusions}
\label{sec:conclusions}
In this work, we have studied the non-relativistic potentials generated for a variety of interactions between spin-1/2 particles. These include interactions mediated by scalars or vectors as well as four fermion operators. We reviewed the procedure laid out in~\cite{Agrawal:2020lea} for setting boundary conditions. In addition to setting the boundary conditions for the Schr\"odinger equation, this procedure also provided us with an analytic cross-check for determining whether a potential is singular or not. Using this diagnostic, we showed that all of the potentials generated from tree-level QFT descriptions of scattering processes give rise to nonsingular well-behaved quantum mechanical potentials. In many cases, this nonsingular behavior is preserved by nontrivial cancellations that arise due to the accompanying operators. Furthermore, we extended this analysis to higher dimensions and showed that Coulomb potentials in an arbitrary number of dimensions are also nonsingular. These results lend credence to the notion of the Quantum Mechanics Swampland and indicate that subject to our matching procedure, singular potentials in quantum mechanics are irrelevant when considering a matching to an underlying perturbative tree-level QFT. We again emphasize that the Quantum Mechanics Swampland is distinct from the Swampland for QFTs, but entirely analogous in its attempt to narrow down the space of consistent, viable quantum mechanical potentials.

We emphasize that all of the evidence presented so far is from tree-level examples. This motivates various different avenues to follow up on our results. In particular, it would be interesting to understand how to extend the matching procedure beyond tree-level. This extension will be relevant for computing loop-level corrections to these potentials as well as processes where the leading order scattering occurs at 1-loop in QFT~\cite{Casmir:1947hx,Feynman:1963ax,Feinberg:1968zz,Iwasaki:1971vb,Feinberg:1989a,Hsu:1992tg,Fischbach:1996qf,Grifols:1996fk,Ferrer:1998ue,BjerrumBohr:2002kt,Fichet:2017bng,Brax:2017xho,Stadnik:2017yge,Thien:2019ayp,Costantino:2019ixl,Ghosh:2019dmi,Segarra:2020rah,Costantino:2020bei,Bolton:2020xsm}. QFTs that spontaneously break Lorentz invariance also yield long-range potentials and it will be interesting to explore the structure of these potentials in more detail~\cite{ArkaniHamed:2004ar}. Nonperturbative QFT, such as coupling to CFT sectors~\cite{Costantino:2019ixl}, presents an additional class of nontrivial examples and at present, it is not clear whether this can give rise to singular potentials from the quantum mechanics viewpoint. As such, it is worth exploring the types of potentials which can arise in these theories. These various investigations will help further refine the boundary between the Swampland and Landscape of quantum mechanical potentials.

\begin{acknowledgments}
We thank Sruthi A. Narayanan, Matthew Reece, and Weishuang Linda Xu for useful discussions and feedback on this manuscript. We also thank Hongwan Liu and Jesse Thaler for useful discussions. This work is supported in part by an NSF Graduate Research Fellowship Grant DGE1745303, the DOE Grant DE-SC0013607, and the Alfred P. Sloan Foundation Grant No.~G-2019-12504.
\end{acknowledgments}

\appendix
\section{Higher Derivative Terms and the Potential Involving Scalars}
\label{sec:scalars}
Scalars do not possess any intrinsic spin. As a result, the non-relativistic limit of the amplitude can only depend on the vector $q^{i}$. In particular, the amplitude, and hence $\Tilde{V}(\vec{q})$, can be parameterized in the following manner
\begin{equation}
    \Tilde{V}(\vec{q}) = \frac{f(q^{2})}{q^{2} + m^{2}} = \sum_{n=0}^{\infty}\frac{a_{n}q^{2n}}{q^{2} + m^{2}}
\label{eqn:scalar_amplitude}
\end{equation}
$m$ is the mass of the mediator and $f(q^{2})$ is a function that can be determined by the specific structure of the interactions in the QFT. We will show explicitly that the first few terms in this series are nonsingular and then generalize to the case of arbitrary $n$.

For concreteness, we will consider an example where $f(q^{2}) = a_{0} + a_{1}q^{2} + a_{2}q^{4}$. We find that
\begin{equation}
\begin{split}
    \Tilde{V}(\vec{q}) &= \frac{a_{0} + a_{1}q^{2} + a_{2}q^{4}}{q^{2} + m^{2}} \\
    &= a_{2}q^{2} + a_{1} - a_{2}m^{2} + \frac{a_{0} - a_{1}m^{2} + a_{2}m^{4}}{q^{2} + m^{2}}
\end{split}
\end{equation}
The last term generates a Yukawa potential and the constant terms generate $\delta^{3}(\vec{r})$. Both of these potentials are nonsingular. The first term is nontrivial. It generates a term proportional to $\nabla^{2}\delta^{3}(\vec{r})$. To show that the first Born approximation in Equation~\ref{eqn:tan_delta_l} is finite, we will evaluate the integrand
\begin{equation}
\begin{split}
    \nabla^{2}\delta(r)r^{\ell+\ell'} &= \frac{1}{r}\frac{\partial^{2}}{\partial r^{2}}\Bigg(\delta(r)r^{\ell+\ell'+1}\Bigg) \\
    &= (\ell+\ell')(\ell+\ell'-1)\delta(r)r^{\ell+\ell'-2}
\end{split}
\end{equation}
The integral is divergent when $\ell + \ell' < 2$, but we see that the coefficient vanishes for those choices of $\ell$ and $\ell'$. We can generalize these results to higher order $q^{2}$ terms. The expression in Equation~\ref{eqn:scalar_amplitude} can be rewritten as follows
\begin{equation}
    \Tilde{V}(\vec{q}) = \sum_{n=0}^{\infty}\frac{a_{n}q^{2n}}{q^{2} + m^{2}} = \frac{\Tilde{a}_{-1}}{q^{2} + m^{2}} + \sum_{n=0}^{\infty}\Tilde{a}_{n}q^{2n}
\end{equation}
The $q^{2n}$ terms generate terms in the potential proportional to $\nabla^{2n}\delta^{3}(\vec{r})$. Evaluating the integrand, we find
\begin{equation}
    \nabla^{2n}\delta(r)r^{\ell+\ell'} = (\ell+\ell')(\ell+\ell'-1)\dotsm(\ell+\ell'+1-2n)\delta(r)r^{\ell+\ell'-2n}
    \label{eqn:higher_derivatives_on_delta_functions}
\end{equation}
This integral is divergent for $\ell + \ell' < 2n$, which happens to be where the coefficient vanishes. Therefore, all the higher derivative terms are well-behaved and produce finite nonsingular first Born approximations.

We can construct a similar argument for the scalar-fermion case as well. Due to the fermion's spin, $q\cdot S$ exists as an additional independent operator. Therefore, the general amplitude can be parameterized as follows
\begin{equation}
	\Tilde{V}(\vec{q}) = \frac{f(q^{2})(1 + \alpha q\cdot S)}{q^{2} + m^{2}} = \sum_{n=0}^{\infty}\frac{a_{n}q^{2n}}{q^{2} + m^{2}} + \frac{b_{n}q^{2n}q\cdot S}{q^{2} + m^{2}}
\end{equation}
The second set of terms in this sum generate terms proportional to $S\cdot\hat{r}\partial_{r}\nabla^{2}\delta^{3}(\vec{r})$. Evaluating the integrand, and using the results of Equation~\ref{eqn:higher_derivatives_on_delta_functions}, we find
\begin{widetext}
\begin{equation}
\begin{split}
S\cdot\hat{r}\partial_{r}\nabla^{2n}\delta(r)r^{\ell+\ell'} &= S\cdot\hat{r}(\ell+\ell')(\ell+\ell'-1)\dotsm(\ell+\ell'+1-2n)\partial_{r}\delta(r)r^{\ell+\ell'-2n} \\
&= S\cdot\hat{r}(\ell+\ell')(\ell+\ell'-1)\dotsm(\ell+\ell'+1-2n)(\ell + \ell' -2n -1)\delta(r)r^{\ell + \ell' - 2n -1}
\end{split}
\end{equation}
\end{widetext}
This integral is divergent for $\ell + \ell' < 2n + 1$. The coefficient vanishes for $\ell + \ell' < 2n$ as before. $\ell + \ell' = 2n$ seems problematic, but here we note that an additional operator exists for this potential. In particular, $S\cdot\hat{r}$ links states with angular momenta that differ by one unit. Therefore, $\ell + \ell'$ must be odd, but $2n$ is manifestly even, and the operator prevents the singularity from arising for this combination of states. 

\section{Evaluating the derivatives on $\delta^{3}(\vec{r})$}
\label{sec:delta_fcn_derivatives}
As an example, we explicitly evaluate the derivatives on the following position space potential:
\begin{equation}
    \frac{\lambda}{m_{1}m_{2}\Lambda^{2}}(\vec{S}_{1}\cdot\vec{\nabla})(\vec{S}_{2}\cdot\vec{\nabla})\delta^{3}(\vec{r})
\end{equation}
We make use of the following relations
\begin{equation}
    \delta^{3}(\vec{r}) = \frac{\delta(r)}{4\pi r^{2}}, \quad -r\delta'(r) = \delta(r), \quad r^{2}\delta''(r) = 2\delta(r)
\end{equation}
In particular, we want to show that the first Born approximation in Equation~\ref{eqn:tan_delta_l} is finite. For clarity of notation, we will omit overall factors. 

\begin{equation}
\begin{split}
\int_{0}^{a}dr s_{\ell'}(kr)(\vec{S}_{1}\cdot\vec{\nabla})(\vec{S}_{2}\cdot\vec{\nabla})\delta^{3}(\vec{r})s_{\ell}(kr) \approx \\
S_{1}^{i}S_{2}^{j}\int_{0}^{a}dr\nabla_{i}\nabla_{j}\frac{\delta(r)}{r^{2}}(kr)^{\ell+1}(kr)^{\ell'+1}
\end{split}
\end{equation}
Now we isolate and evaluate the integrand.

\begin{widetext}
\begin{equation}
    \begin{split}
        \nabla_{i}\nabla_{j}\delta(r)r^{\ell+\ell'} &= \delta(r)\partial_{i}\partial_{j}r^{\ell+\ell'} + \partial_{i}r^{\ell+\ell'}\partial_{j}\delta(r) + \partial_{j}r^{\ell+\ell'}\partial_{i}\delta(r) + r^{\ell+\ell'}\partial_{i}\partial_{j}\delta(r) \\
        &= \delta(r)r^{\ell+\ell'-2}\Bigg[(\ell+\ell'-1)\delta_{ij} + (3 + (\ell+\ell')(\ell+\ell'-4))\hat{r}_{i}\hat{r}_{j}\Bigg]
    \end{split}
\end{equation}
\end{widetext}
%\iffalse
%\begin{equation}
%    \begin{split}
%        \nabla_{i}\nabla_{j}\delta(r)r^{\ell+\ell'} &= \delta(r)\partial_{i}\partial_{j}r^{\ell+\ell'} + \partial_{i}r^{\ell+\ell'}\partial_{j}\delta(r) + \partial_{j}r^{\ell+\ell'}\partial_{i}\delta(r) + r^{\ell+\ell'}\partial_{i}\partial_{j}\delta(r) \\
%        &= \delta(r)\partial_{i}\frac{r_{j}}{r}(\ell+\ell')r^{\ell+\ell'-1} + 2(\ell+\ell')r^{\ell+\ell'-1}\frac{r^{i}}{r}\frac{r^{j}}{r}\delta'(r) + r^{\ell+\ell'}\partial_{i}\frac{r_{j}}{r}\delta'(r) \\
%        &= \delta(r)(\ell+\ell')\Bigg[r^{\ell+\ell'-2}\delta_{ij} - r_{j}\frac{r_{i}}{r}(\ell+\ell'-2)r^{\ell+\ell'-3}\Bigg] 2(\ell+\ell')r^{\ell+\ell'-1}\frac{r^{i}}{r}\frac{r^{j}}{r}\delta'(r) + r^{\ell+\ell'}\Bigg[\delta_{ij}\frac{\delta'(r)}{r} + r_{j}\frac{r_{i}}{r}\frac{r\delta''(r) - \delta'(r)}{r^{2}}\Bigg] \\
%        &= \delta(r)r^{\ell+\ell'-2}\Bigg[(\ell+\ell'-1)\delta_{ij} + (3 + (\ell+\ell')(\ell+\ell'-4))\hat{r}_{i}\hat{r}_{j}\Bigg]
%    \end{split}
%\end{equation}
%\fi
We see that the integral is divergent when $\ell + \ell' < 2$. If $\ell = \ell' = 0$, the operator simplifies to $\mathcal{O}_{T}$ which vanishes when sandwiched between states of $\ell = \ell' = 0$. If $\ell = 1$ or $\ell' = 1$, then the integrand also vanishes. The cancellation is nontrivial in both cases and the operator structure conspires to produce a nonsingular first Born approximation. We have checked explicitly that the other potentials in Section~\ref{sec:nonrel_ops} also produce nonsingular first Born approximations.

\bibliography{ref.bib}

\end{document}